\documentclass[twocolumn,superscriptaddress,showpacs,floatfix,preprintnumbers, nofootinbib,hyperref]{revtex4} 
\usepackage{graphicx}%  Default Latex 2eps package for embedding figures
\usepackage{rotating}
\usepackage{epsfig}
\usepackage{bm}
\usepackage[usenames,dvipsnames]{xcolor}
\usepackage{color}

\def\he4{$^4$He}
\def\h2{$^2$H}

\newcommand{\lesssim}{\,\rlap{\lower3.7pt\hbox{$\mathchar\sim$}}
\raise1pt\hbox{$<$}\,}

\begin{document}

\preprint{ IPPP/14/13, DCPT/14/26, MPP-2014-21}

\title{{Suppression of the multi-azimuthal-angle instability
in dense neutrino gas \\ during supernova accretion phase
}}

\author{Sovan Chakraborty}
\affiliation{Max-Planck-Institut f\"ur Physik
(Werner-Heisenberg-Institut)\\
 F\"ohringer Ring 6, D-80805 M\"unchen, Germany}

\author{Alessandro Mirizzi} 
\affiliation{II Institut f\"ur Theoretische Physik, Universit\"at Hamburg, Luruper Chaussee 149, 22761 Hamburg, Germany}

\author{Ninetta Saviano} 
\affiliation{Institute for Particle Physics Phenomenology, Department of Physics,
Durham University,\\ Durham DH1 3LE, United Kingdom}

 \author{David de Sousa   Seixas}
\affiliation{Max-Planck-Institut f\"ur Physik
(Werner-Heisenberg-Institut)\\
 F\"ohringer Ring 6, D-80805 M\"unchen, Germany}

%\date{\today}

\begin{abstract}
It has been recently pointed out that removing the axial symmetry in the  ``multi-angle effects'' 
associated with the neutrino-neutrino interactions for supernova (SN) neutrinos, a new multi-azimuthal-angle (MAA) instability
would arise. In particular, 
for a flux ordering $F_{\nu_e} > F_{\bar\nu_e} > F_{\nu_x}$, as expected during the SN accretion phase, this instability  occurs  in the normal neutrino mass hierarchy. 
However, during this phase the ordinary matter density can be larger than the neutrino one, suppressing the self-induced conversions. At this regard, 
 we investigate the matter suppression
of the MAA effects, performing a linearized stability analysis of the neutrino equations of motion, in the presence of realistic SN density profiles. We compare these results with the numerical solution of 
the SN neutrino non-linear evolution equations.
Assuming axially symmetric distributions of neutrino momenta 
we find that the large matter term strongly inhibits the MAA effects. In particular,  the hindrance becomes
 stronger including 
realistic forward-peaked neutrino angular distributions. 
As a result, in our model for a $10.8$~$M_{\odot}$ iron-core SNe, MAA instability does not trigger any flavor conversion during the 
accretion phase. Instead, for a  $8.8$~$M_{\odot}$  O-Ne-Mg core SN model, with lower matter density profile and less forward-peaked angular distributions, flavor conversions are possible also at early times.

\end{abstract}


\maketitle

\section{Introduction}

The  flavor evolution of supernova (SN) neutrinos, 
is strongly impacted by the self-induced effects, 
associated with instabilities induced by   the  the neutrino-neutrino
interactions in the deepest stellar
 regions~\cite{Pantaleone:1992eq,Qian:1994wh,Sawyer:2005jk,Duan:2005cp,Duan:2006an,Hannestad:2006nj,Fogli:2007bk,Fogli:2008pt}
(see also~\cite{Duan:2010bg} for a recent review).
In this context a key ingredient  in the characterization of these  effects  is related to
the  current-current nature of low-energy weak interactions, which implies 
a ``multi-angle term''~\cite{Duan:2006an,Sawyer:2008zs} $(1-{\bf v}_{\bf p} \cdot {\bf v}_{\bf q})$, where ${\bf v}_{\bf p}$
is the neutrino velocity~\cite{Qian:1994wh}. 
Till recently,  all   studies  have assumed the 
axial symmetry for this term, that would then depend only on the zenith-angle. 
It has been shown that the ``multi-zenith-angle'' (MZA) term
 in some cases can hinder the maintenance    of the coherent oscillation 
behavior for different neutrino 
modes~\cite{Duan:2006an,Raffelt:2007yz,Fogli:2007bk,EstebanPretel:2007ec,Sawyer:2008zs,Mirizzi:2010uz}.  
 
 A valuable tool to diagnose the possible instabilities of a dense neutrino gas
is given by  the linearized stability analysis of the neutrino equations of motion.
The linearized equations including generic azimuthal and zenith angle distributions for neutrinos
were first worked out  in~\cite{Sawyer:2008zs}. Then, the formalism for the azimuthal symmetric case
was further developed in~\cite{Banerjee:2011fj}.
 The stability method would  allow one to determine  the possible onset of the flavor conversions, seeking  for
an exponentially growing solution of the eigenvalue problem, associated with the linearized equations of motion for
the neutrino ensemble. 
Recently, this method has been applied in~\cite{Raffelt:2013rqa}
 removing  the  assumption of axial symmetry in the $\nu$ propagation.
 As a result, a   multi-azimuthal-angle (MAA) instability was found
in addition to the  bimodal~\cite{Samuel:1995ri} and MZA ones~\cite{Raffelt:2013rqa}.
 In particular, it was considered a  neutrino ensemble with a strong excess of 
$\nu_e$ over $\bar\nu_e$, as expected during the SN accretion phase (at post-bounce times $t_{\rm pb} \lesssim 0.5$~s).
In this situation, the instability
has been found in normal mass hierarchy (NH,  ${\Delta m^2_{\rm atm}} = m_3^2-m_{1,2}^2>0$), where the system
would have been stable imposing   the  perfect axial symmetry. 
 Subsequently, the role of this instability   has been clarified in~\cite{Raffelt:2013isa,Duan:2013kba}
 with simple toy models. 
 The discovery of the new MAA effects, 
 has also motivated first numerical studies of the non-linear 
 neutrino propagation equations in SN, 
introducing
the azimuthal angle as angular variable in the multi-angle kernel,  in addition to the usual zenith 
angle~\cite{Mirizzi:2013rla,Mirizzi:2013wda}.
Remarkably it was considered  the $\nu$ propagation  only along a radial direction, i.e.  a local solution  along a specific line of sight, under the assumption that the transverse variations of the global solution are small.
In this approximation, for the  unstable case discussed above, MAA effects   lead in NH
to spectral swaps and splits analogous  to what produced in inverted hierarchy (IH,  ${\Delta m^2_{\rm atm}}<0$)
 by the known bimodal instability~\cite{Samuel:1995ri} also in a completely isotropic neutrino gas~\cite{Raffelt:2007cb}.

All these recent works assume that the MAA effects
 can develop without any matter hindrance. 
Remarkably,  a crucial  ingredient  to be considered to determine the 
impact of self-induced flavor instabilities is the  ordinary matter term, associated with the net
electron densities $n_e$ in SNe.
As pointed
out in~\cite{EstebanPretel:2008ni} for the axial-symmetric case, when $n_e$
is not negligible with respect to
the neutrino density $n_\nu$, the large phase dispersion induced by the matter for
$\nu$'s traveling in different directions, would partially or
totally suppress the self-induced oscillations through peculiar MZA effects. 
Recent studies of this case performed with realistic SN models, indicates that
this situation is  realized during the supernova accretion phase
(at post-bounce times $t_{\rm pb} \lesssim 0.5$~s). As a consequence, the self-induced
flavor conversions found in IH in the axial-symmetric models are strongly 
inhibited~\cite{Chakraborty:2011nf,Chakraborty:2011gd,Saviano:2012yh,Sarikas:2011am}.

From a preliminary schematic study done in~\cite{Raffelt:2013rqa} it results that the 
matter density required to suppress the MMA instability in NH   is larger than the one necessary
to suppress the self-induced conversions in IH for the axial symmetric case.  
Motivated by these previous results, we find it is  mandatory to understand  what is the role
of the dense matter on the MAA effects during the accretion phase. 
We will use   as benchmark for the neutrino  and  matter density profiles the SN models 
from  recent long term simulations of core-collapse  explosions, performed by the
Basel-Darmstadt model. These were already considered by some of us in~\cite{Chakraborty:2011nf,Chakraborty:2011gd,Saviano:2012yh}.  
The plan of our work is as follows.
In Sec.~2  we present the neutrino equations of flavor evolution  without 
imposing axial asymmetry. Then, we describe  the setup to perform the  stability 
analysis of the linearized equations of motion.   
In Sec.~3 we present the results of the stability analysis and we compare them with
the numerical solution of the equations. We focus on the accretion phase for two different
 SN progenitor models. Finally, in Sec.~4 we comment on our results and we conclude.

\section{Setup of the flavor evolution}

\subsection{Equations of flavor evolution}
 
 In the axial symmetric case, the SN neutrino flavor evolution is described 
 by ordinary differential equations~\cite{Sigl:1992fn}, characterizing the flavor changes along a radial 
direction.  
 When axial symmetry is broken by the MAA effects,   in order to get   the global solution of the problem
in general one would  consider also variations along the  transverse
direction to the neutrino propagation. This would imply passing from  ordinary to  partial differential equations, with a big
layer of complication in the numerical solution.  
However, 
in our study we are mostly interested in answering the question of the  stability
of the dense neutrino gas under MMA effects, in the presence  of a large matter term. 
Therefore, \emph{before} the MAA instability emerges, we can still consider the ordinary differential equations in the only radial direction. 
These are enough to determine which   cases are completely stable under  MAA effects. In the other cases, 
in which the MAA instability  develops, these equations would  be useful to determine
the onset radius of the flavor conversions. 
Nevertheless, the subsequent flavor evolution can be taken just as indicative, since it is based  
 on the assumption that   variations in the transverse direction
always remain small.  

Under this approximation,  the flavor evolution  depends only on $r$, $E$ and ${\bf v}_{\bf p}$.
 Then, following~\cite{Banerjee:2011fj, Raffelt:2013rqa} we write the equations of motion for the flux matrices  $\Phi_{E,u, \varphi}$  as function of the radial coordinate.
 We use negative energy $E$  for anti-neutrinos. 
Following the usual prescription, we label the zenith angular mode in terms of the variable 
$u=\sin^2\theta_R$, where
  $\theta_R$ is the emission angle relative to the radial direction
of the neutrinosphere radius $R_{\nu}$~\cite{EstebanPretel:2007ec,Banerjee:2011fj}.
We call  $\varphi$  the azimuth angle of the neutrino velocity ${\bf v_p}$. 
We normalize the flux matrices to the  ${\overline\nu}$ number flux at the neutrinosphere.
The diagonal $\Phi_{E,u, \varphi}$ elements are
the ordinary number fluxes 
integrated
over a sphere of radius $r$.
 The off-diagonal elements,
which are initially zero, carry a phase information due to 
flavor mixing.
Then, the equations of motion read~\cite{Banerjee:2011fj,Raffelt:2013rqa}
%.....................................................
\begin{equation}
\textrm{i}\partial_r \Phi_{E,u, \varphi}=[H_{E,u, \varphi},\Phi_{E,u,\varphi}] \,\ ,
\label{eq:eom1}
\end{equation}
%......................................................... 
with the Hamiltonian~\cite{Raffelt:2013rqa}
%.......................................................
 \begin{eqnarray}
& & H_{E,u} = \frac{1}{v_{u}} \left(\frac{M^2}{2E} + \sqrt{2} G_F N_l \right) \nonumber \\
 &+& \frac{\sqrt{2}G_F}{4\pi r^2} \int d \Gamma_{E,u, \varphi}^{\prime}  \left(\frac{1-v_{u}v_{u^\prime}-{\bm\beta}\cdot {\bm\beta}^{\prime}}
{v_{u}v_{u^\prime}} \right)  \Phi^\prime \,\ .
 \label{eq:eom2}
 \end{eqnarray}
%......................................................... 
The matrix $M^2$ of neutrino mass-squares causes vacuum
flavor oscillations. 
We work in a two-flavor scenario, associated with 
the atmospheric mass-square difference $\Delta m^2_{\rm atm}= 2 \times 10^{-3}$~eV$^2$
and  a small (matter suppressed) in-medium mixing $\Theta = 10^{-3}$. We will always consider NH, where MAA effect could emerge
for the flux ordering we are considering.
The matrix $N_l= \textrm{diag}(n_e,0,0)$ in flavor basis, contains the net electron density and 
is responsible for the Mikheyev-Smirnov-Wolfenstein (MSW) matter effect~\cite{Matt}   with the ordinary background. 
Finally, 
 the term at second line represents the $\nu$-$\nu$ refractive 
term, where $\int d \Gamma_{E,u, \varphi} = \int_{-\infty}^{+\infty}d E
 \int_{0}^{1}du  \int_{0}^{2\pi}d\varphi$.
 In the multi-angle term of Eq.~(\ref{eq:eom2}),
 the radial velocity of a mode with angular label $u$ is
$v_{u} = (1-u R_{\nu}^2/r^2)^{1/2}$~\cite{EstebanPretel:2007ec} and the transverse velocity is 
$\beta_{u}= u^{1/2} R_{\nu}/r$~\cite{Raffelt:2013rqa}.
The term 
$ {\bm\beta}\cdot{\bm\beta}^{\prime} = \sqrt{u u^{\prime}}{R_{\nu}^2/r^2}
\cos(\varphi-\varphi^{\prime})$ is the 
responsible for the breaking of the axial symmetry. 

To solve numerically  Eq.~(\ref{eq:eom1})  we use an integration routine for stiff ordinary differential equations taken
from the NAG libraries~\cite{nag} and based on an adaptive method.
We have used  $N_\varphi=30$ modes for $\varphi \in [0;2\pi]$, $N_u=1400$
for $u \in [0;1]$.
Concerning neutrino emission model, in order to simplify the complexity of our numerical simulations of the flavor evolution, we assume all
$\nu$'s to be represented by a single energy, that we fix at
$E=15$~MeV. This
approximation is reasonable since our main purpose is to
determine only if the dense matter effects block the development of
the self-induced transformations.

 \subsection{Stability conditions}

In order to perform the stability analysis, we  linearize the equations of motion
[Eq.~(\ref{eq:eom2})],  following the approach of~\cite{Banerjee:2011fj,Raffelt:2013rqa}. 
We   write the flux matrices in the form
%...........................................................
\begin{equation}
\Phi_{\omega,u}= \frac{\textrm{Tr}\Phi_{\omega,u, \varphi}}{2}+
\frac{g_{\omega,u, \varphi}}{2}
\left( \begin{array}{cc} s &  S \\
S^{\ast} & -s
\end{array} \right) \,\ ,
\label{eq:pHI}
\end{equation}
%..............................................................
where we switch to the 
frequency variable $\omega= \Delta m^2_{\rm atm}/2E$, and  we introduce the 
neutrino flux difference distributions $g_{\omega,u, \varphi}\equiv g(\omega,u, \varphi)$ 
that represent the flavor fluxes $n_{\nu_e}(R_{\nu})-n_{\nu_x}(R_{\nu})$ at the neutrinosphere, 
normalized  to the ${\overline\nu}$ flux. 
In the following we will always assume axial symmetry of the neutrino emission. Therefore $ g(\omega,u, \varphi)=  g(\omega,u)/2\pi$.
The $\textrm{Tr}\Phi_{\omega,u, \varphi}$ is conserved and then irrelevant for the flavor conversions.
The $\nu_e$ survival probability is $\frac{1}{2}(1+s)$, given in terms of the swap factor $-1 \leq s \leq 1$
of the  
 matrix in the second term on the right-hand side.
The off-diagonal components  $S$ are complex and $s^2 + |S|^2=1$. 
The initial conditions are $s=1$ and $S=0$.
Self-induced flavor transitions  start when the off-diagonal term $S$ grows
 exponentially.

In the small-amplitude limit $|S|\ll 1$, 
and at far distances from the neutrinosphere $r \gg R_{\nu}$,
the linearized evolution equations for  $S$
assume the form~\cite{Raffelt:2013rqa}
%..................................................
\begin{eqnarray}
\textrm{i}\partial_r S &=& [\omega + u(\lambda +\epsilon \mu)] S  \nonumber \\
&-& \mu \int  d \Gamma^\prime [u + u^\prime - 2\sqrt{u u^\prime} \cos (\varphi-\varphi^\prime)]g^\prime S^\prime
\label{eq:lin} \,\ .
\end{eqnarray}
%......................................................
%...........................................................
In this equation
$
\epsilon =  \int d \Gamma_{\omega,u,\varphi}\,\ g_{\omega,u,\varphi} 
$,
quantifies the ``asymmetry'' of the neutrino spectrum, normalized to the  ${\overline\nu}$ 
number flux. In  the SN models we are using, typically $\epsilon \sim 0.3-0.5$ during the accretion phase
(see  Fig. 3 in~\cite{Chakraborty:2011gd}).

The $\nu$-$\nu$ interaction strength is given by
%.................................................................
\begin{eqnarray}
\mu &=& \frac{\sqrt{2}G_F [n_{{\bar\nu}_e}(R_{\nu})-n_{{\bar\nu}_x}(R_{\nu})]}{4 \pi r^2}\frac{R_{\nu}^2}{2 r^2} \nonumber \\
&=& \frac{3.5 \times 10^{9}}{r^4}
\left(\frac{L_{{\overline\nu}_e}}{\langle E_{{\overline\nu}_e} \rangle}
- \frac{L_{{\overline\nu}_x}}{\langle E_{{\overline\nu}_x} \rangle} \right)
\nonumber \\
& \times &
% \left(\frac{L_{{\overline\nu}_e}}{10^{52} \,\ \textrm{erg/s}}\right)
\left(\frac{15 \,\ \textrm{MeV}}{
10^{52} \,\ \textrm{erg/s}}\right) 
 \left(\frac{R_{\nu}}{10 \,\ \textrm{km}} \right)^2 ,  
\label{eq:mu}
\end{eqnarray}
%..................................................................
while  ordinary matter background term is given by 
%............................................................
\begin{eqnarray}
\lambda &=& \sqrt{2} G_F n_e \frac{R_{\nu}^2}{2 r^2} \nonumber \\
&=& \frac{0.95 \times 10^8 }{r^2} \left(\frac{Y_e}{0.5} \right) 
\left(\frac{\rho}{10^{10} \textrm{g}/\textrm{cm}^3} \right) 
\left(\frac{R_{\nu}}{10 \,\ \textrm{km}} \right)^2   
\label{eq:lambda}
\end{eqnarray}
%.............................................................
where $Y_e$ is the net electron fraction,  and $\rho$ is the matter density. 
The radial distance $r$ is expressed in km, 
while the numerical values of $\mu$ and  $\lambda$ in the two previous equations 
are quoted in km$^{-1}$, as appropriate for the SN case.

One can write the solution of the linear differential equation [Eq.~(\ref{eq:lin})]
in the form $S = Q _{\omega,u, \varphi} e^{-i\Omega r}$ with complex frequency
$\Omega= \gamma + i \kappa$ and eigenvector $Q _{\omega,u, \varphi}$. A solution with 
$\kappa >0$ would indicate an exponential increase in $S$, i.e. an instability.
The solution of Eq.~(\ref{eq:lin}) can then be recast in the form of an eigenvalue equation
for $Q _{\omega,u, \varphi}$. One gets as consistency condition~\cite{Raffelt:2013rqa}
%%%%%%%%%%%%%%%%%%%%%%%%%%%%%%%%%%%%%%%%%%%%%%%%%%%%%%%5
\begin{equation}
\mu \int d\omega du \,\  \frac{ u g_{\omega,u}}
{\omega +u(\lambda +\epsilon \mu) -\Omega} +1=0 \,\ .
\label{consit}
\end{equation}
%%%%%%%%%%%%%%%%%%%%%%%%%%%%%%%%%%%%%%%%%%%%%%%%%%%%%%%%%%

A flavor instability is present whenever Eq.~(\ref{consit}) admits a  solution $(\gamma, \kappa)$. 

%%%%%%%%%%%%%%%%%%%%%%%%%%%%%%%%%%%%%%%%%%%%%%%%%%%%%%%%%%%%%%%%%%%%
\section{APPLICATION TO OUR SUPERNOVA
MODELS}
%%%%%%%%%%%%%%%%%%%%%%%%%%%%%%%%%%%%%%%%%%%%%%%%%%%%%%%%%%%%%%%%%%%%%%

We consider the core-collapse supernova simulations
of massive  stars with 8.8 and 10.8~$M_{\odot}$
progenitor from Ref.~\cite{Fischer:2009af}, taken as benchmark for our numerical
study in~\cite{Chakraborty:2011nf,Chakraborty:2011gd,Saviano:2012yh}. The first type of SN belongs to the class of O-Ne-Mg-core progenitor. The second one is an iron-core progenitor.

Under the single-energy  approximation we are using, we characterize  the neutrino energy spectra as 
%..........................................................................................   
\begin{eqnarray}
g(\omega, u) &=&\frac{1}{n_{\bar\nu_e}-n_{\bar\nu_x}}[
(n_{\nu_e} g_{\nu_e}(u)-n_{\nu_x} g_{\nu_x}(u))\delta(\omega-\omega_0) \nonumber \\
&-&
(n_{\bar\nu_e}g_{\bar\nu_e}(u)-n_{\bar\nu_x}g_{\bar\nu_x}(u))\delta(\omega+\omega_0)] \,\ ,
\label{singleen}  
\end{eqnarray}
%.........................................................................................
where $n_{\nu_\alpha}$ are the total number fluxes of the species $\nu_{\alpha}$ at the neutrinosphere.
In order to fix the neutrinosphere radius $r=R_{\nu}$, consistently with our choice 
in~\cite{Chakraborty:2011nf,Chakraborty:2011gd,Saviano:2012yh}  we take the radius at which the
$\nu_e$'s angular distribution has no longer significant backward
flux, i.e. a few \% of the total one. 
This typically is in
the range
$R \sim 50-100$~km (see Fig.~4 in~\cite{Chakraborty:2011gd}).

For our choice of neutrino representative energy ($E=15$~MeV), the corresponding frequency is 
%..............................................................................
\begin{equation}
\omega_0 = \left\langle \frac{\Delta m_{\rm atm}^2}{2E} \right\rangle = 0.34 \,\ \textrm{km}^{-1} \,\ .
\end{equation}
%................................................................................
The $g_{\nu_\alpha}(u)$ represent the zenith-angle distributions. 
In our  study we assume two different models. At first, we consider the 
``half-isotropic'' case, where in analogy with  a black-body emission, it is assumed
$g_{\nu_\alpha}(u)=1$ for all the species.
We will then compare the results obtained in this widely used prescription, with the one obtained
taking the $g_{\nu_\alpha}(u)$ directly from the output of the SN simulations. 
In this case, the zenith-angle distributions would be flavor-dependent and  forward enhanced (i.e. peaked at small
$u$)
with respect to the half-isotropic emission model  (see Fig.~1 and the 
discussion in~\cite{Saviano:2012yh}). 
 We will see  that the presence of forward peaked distributions will enhance the matter suppression of the MAA instability. 
Finally, we comment that our results are based on axial symmetric neutrino angular distributions. 
In this regard, it is interesting to mention that in~\cite{Sawyer:2008zs} it has been shown
that if  perfect cylindrical symmetry in the initial neutrino distributions
were given up,   super-fast flavor turnovers could be produced. We leave the investigation of this
interesting issue for a future work.

%%%%%%%%%%%%%%%%%%%%%%%%%%%%%%%%%%%%%%%%%%%%%%%%%%%%%%%%%%%%%%%%%%%%%%%%%%%%%%%%%%%%%%%%%%%%%%%%%%%%%%%%%%%%%
\begin{figure}[!t]
\begin{center}
 \includegraphics[angle=0,width=0.5\textwidth]{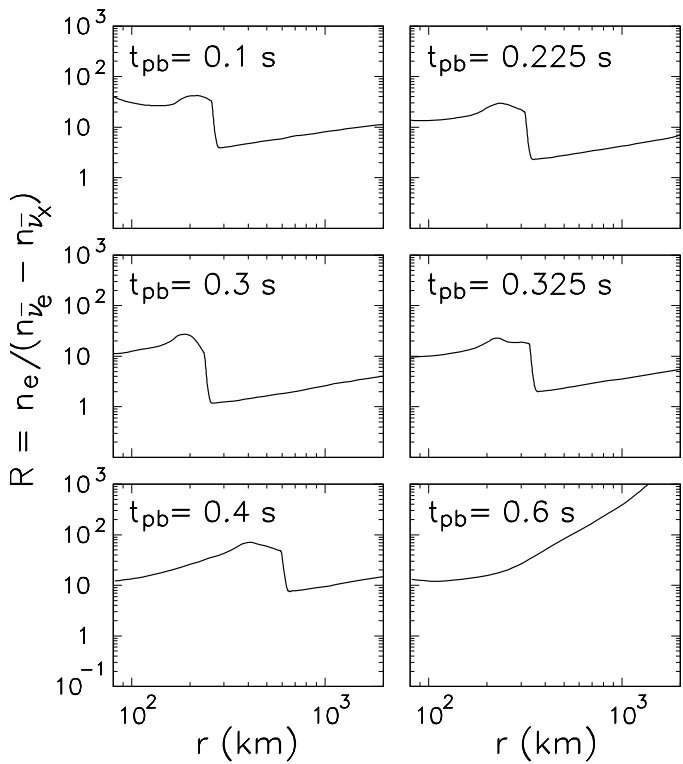} 
    \end{center}
\caption{10.8~$M_{\odot}$
progenitor mass. Radial evolution of the ratio
$R$ between the matter $\lambda$ and neutrino  $\mu$ potentials at different
post-bounce times.} \label{fig1}
\end{figure}
%%%%%%%%%%%%%%%%%%%%%%%%%%%%%%%%%%%%%%%%%%%%%%%%%%%%%%%%%%%%%%%%%%%%%%%%%%%%%%%%%%%%%%%%%%%%%%%%%%%%%%%%%%%%%%%%%%  

\subsection{$10.8$~$M_{\odot}$}

We start our investigation with the case of the $10.8$~$M_{\odot}$
iron-core supernova.
For this  model, 
the net electron density $n_e$ and the neutrino densities $n_\nu$ for different post-bounce times
were shown in Fig.~5 of Ref.~\cite{Chakraborty:2011gd}. 
 In order to quantify the relative strength of 
the matter potential $\lambda$ [Eq.~(\ref{eq:lambda})] with respect to the neutrino potential $\mu$
[Eq.~(\ref{eq:mu})]
 we plot in Fig.~1 the ratio $R=\lambda/\mu$ as a function of the radial coordinate
$r$ at different post-bounce times $t_{\rm pb}$ in the range [0.1,0.6]~s. 
We realize that $R > 10$ before the abrupt discontinuity associated with the shock front position.
As we will see with the stability analysis, this strong matter dominance would prevent the flavor conversions 
before the shock front. 
However, for some time snapshots (i.e. $t_{\rm pb}= 0.225-325$~s) the ratio can become $R \gtrsim 1-2$ after the shock front, leaving the possibility of flavor conversions in this region.

For the
 same post-bounce times 
of Fig.~1,  we show in Fig.~2 the radial evolution of  the eigenvalue $\kappa$ determined 
from the solution of Eq.~(\ref{consit}).
We consider the following cases: \emph{ (a)}   $\lambda=0$ and a half-isotropic neutrino emission (dashed curves),
\emph{ (b)}
 dense matter effects and
   a half-isotropic neutrino emission
(continuous curves) and,  \emph{ (c)} dense  matter effects and non-trivial  neutrino angular distributions 
(dotted curves). 
In Fig.~3 we show the survival probability $P_{ee}$ for electron antineutrinos ${\bar\nu}_e$ for the same cases of Fig.~2, obtained solving
the non-linear propagation equations [Eq.~(\ref{eq:eom2})].

We start discussing our results for the case of $\lambda=0$ [case \emph{(a)}]. 
We realize that when the neutrino system enters  the unstable regime ($\kappa >0$), the $\kappa$
function rapidly grows from zero to a peak value greater than one. Only for $t_{\rm pb}=0.6$~s $\kappa \sim 0.3$
 at the peak. Indeed, for this time the asymmetry parameter $\epsilon >2$ (see  Fig. 3 in~\cite{Chakraborty:2011gd}). Then the consistency condition
Eq.~(\ref{consit}) in order to be satisfied requires a smaller $\kappa$.
Comparing the results of Fig.~2 and 3 one finds a good agreement between the numerical onset of the self-induced
flavor conversions triggered by the MAA effect and the position of the  peak in the $\kappa$ function. 
  
We   now discuss  the situation of   realistic matter density profiles and 
a half-isotropic neutrino emission [case \emph{(b)}]. 
As expected, the flavor instability is strongly suppressed with respect to the previous
case without matter. In particular, the $\kappa$ function, when not completely vanishing (as at $t_{\rm pb}=0.4$~s), 
 would start growing only after the shock front position [see Fig.~1]. This is due to 
the fact that at lower radii the ratio $R \gg 1$. 
The   $\kappa$ function then reaches peak values between 0.5 and 1 only at intermediate times, i.e. $t_{\rm pb}= 0.3, 0.325$~s, for which  the ratio
$R \gtrsim 1$ in the post-shock region. 
For the other time snapshots $\kappa$ it is more suppressed, consistently with a larger value of $R$. 
Comparing these results with the numerical calculation of the $P_{ee}$ in Fig.~3, we realize
 that the presence of a non-zero $\kappa$ is not enough to guarantee the onset of flavor conversions.
 Indeed, for $t=0.1,  0.6$~s, the $\kappa$ function is too small and dies out too quickly
before  triggering 
 flavor conversions. 
 For the cases in which flavor conversions occur, i.e. at  $t_{\rm pb}= 0.225, 0.3, 0.325$~s, the numerical onset is shifted at larger radii
 by few hundred km, with respect to the peak of the $\kappa$ function. This delay is due to the fact that since
 the instability is weaker with respect to the case with $\lambda=0$, 
the slower rate of growth implies
a larger radial distance in order to develop 
 significant effects on the $P_{ee}$. 
 We checked that  a non-zero  $\kappa$  corresponds 
 to the exponential growth of the off-diagonal components $S \sim e^{\kappa  r}$ of the flux matrices [see Eq.~(\ref{eq:pHI})], while
 the  change of the diagonal components, would occur only at larger radii.
It would be interesting to
see if the stability analysis can be further developed in order to achieve a better
understanding of this  dynamics.  

Then we  consider the case in which also the flavor-dependent forward-peaked neutrino angular
distributions are taken into account [case \emph{(c)}]. 
We find that the $\kappa$ function in this case is completely suppressed. 
This is consistent  with the expectation that the $\nu$-$\nu$ strength is 
weaker for forward-peaked distributions, making more effective the matter suppression.
This result is consistent with the output of the numerical simulations  that show 
for all the considered time snapshots $P_{ee}=1$.
 
 Finally we mention that in~\cite{Cherry:2012zw,Sarikas:2012vb} it has been
 claimed that  possible residual scatterings could affect $\nu$'s after the 
neutrinosphere, producing a small ``neutrino halo'' that would broaden the $\nu$ angular 
distributions~\cite{Cherry:2012zw,Sarikas:2012vb} at $r \gtrsim 100$~km. We checked that 
the results of the SN simulations we are using have not enough angular resolution to exhibit this 
feature. However, in order to characterize the possible halo effect, we performed the same analytical estimation
presented in~\cite{Sarikas:2012vb}. We repeated the stability analysis including the halo effect in the $\nu$ angular
distributions, without finding any change with respect to the results shown here. 
Therefore, we conclude that for our   $10.8$~$M_{\odot}$ SN model, MAA instability is always suppressed by the dense matter effects
during the accretion phase.

%%%%%%%%%%%%%%%%%%%%%%%%%%%%%%%%%%%%%%%%%%%%%%%%%%%%%%%%%%%%%%%%%%%%%%%%%%%%%%%%%%%%%%%%%%%%%%%%%%%%%%%%%%%%%
\begin{figure}[!t]
\begin{center}
 \includegraphics[angle=0,width=0.5\textwidth]{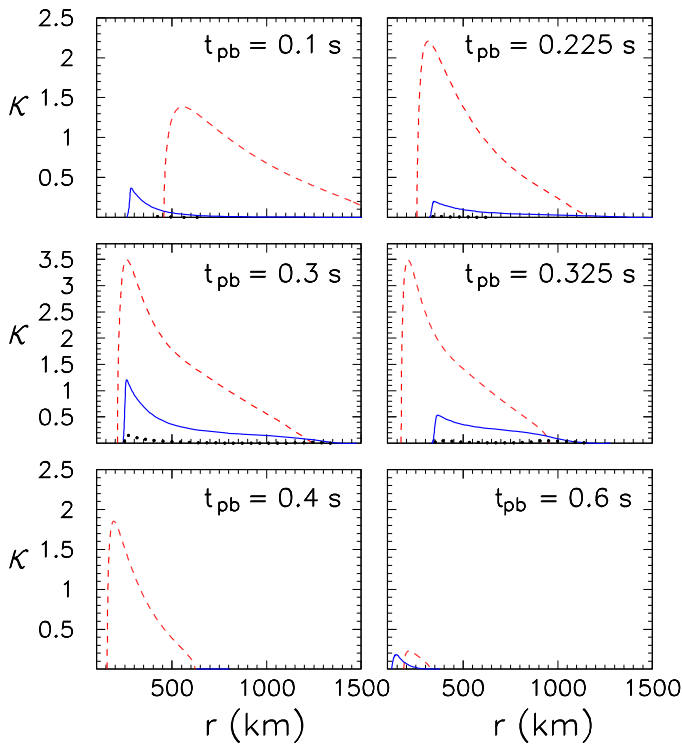} 
    \end{center}
\caption{10.8 M$_{\odot}$ progenitor mass. Radial evolution of the 
$\kappa$ function (in units of km$^{-1}$) at different post-bounce times
 with $\lambda=0$ for a half-isotropic neutrino emission
(dashed curves)  and in presence of matter effects, with 
a half-isotropic neutrino emission (continuous curves) and 
with flavor-dependent angular distributions (dotted curves).} \label{fig2}
\end{figure}
%%%%%%%%%%%%%%%%%%%%%%%%%%%%%%%%%%%%%%%%%%%%%%%%%%%%%%%%%%%%%%%%%%%%%%%%%%%%%%%%%%%%%%%%%%%%%%%%%%%%%%%%%%%%%%%%%%  

%%%%%%%%%%%%%%%%%%%%%%%%%%%%%%%%%%%%%%%%%%%%%%%%%%%%%%%%%%%%%%%%%%%%%%%%%%%%%%%%%%%%%%%%%%%%%%%%%%%%%%%%%%%%%
\begin{figure}[!t]
\begin{center}
 \includegraphics[angle=0,width=0.5\textwidth]{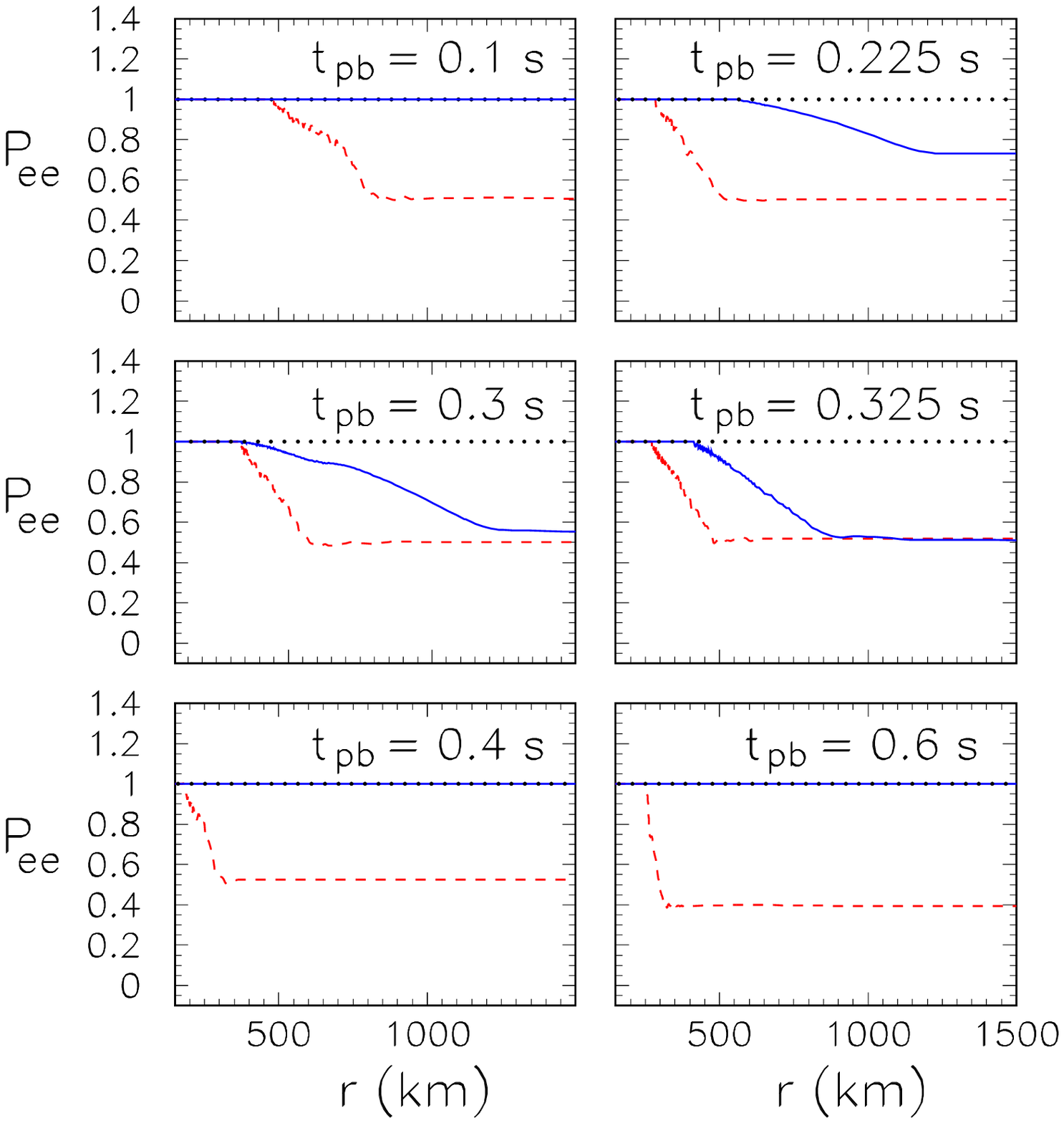} 
    \end{center}
\caption{10.8 M$_{\odot}$ progenitor mass. Radial evolution of the
survival probability $P_{ee}$
for electron antineutrinos at different
post-bounce times for the MAA evolution with $\lambda=0$ for a half-isotropic neutrino emission
(dashed curves)  and in presence of matter effects, with 
a half-isotropic neutrino emission (continuous curves) and 
with flavor-dependent angular distributions (dotted curves).} \label{fig3}
\end{figure}
%%%%%%%%%%%%%%%%%%%%%%%%%%%%%%%%%%%%%%%%%%%%%%%%%%%%%%%%%%%%%%%%%%%%%%%%%%%%%%%%%%%%%%%%%%%%%%%%%%%%%%%%%%%%%%%%%% 

\subsection{$8.8$~$M_{\odot}$}

%%%%%%%%%%%%%%%%%%%%%%%%%%%%%%%%%%%%%%%%%%%%%%%%%%%%%%%%%%%%%%%%%%%%%%%%%%%%%%%%%%%%%%%%%%%%%%%%%%%%%%%%%%%%%
\begin{figure}[!t]
\begin{center}
 \includegraphics[angle=0,width=0.5\textwidth]{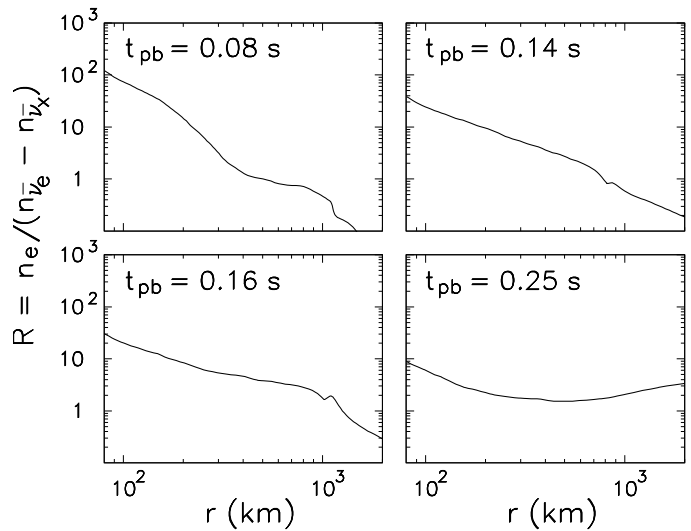} 
    \end{center}
\caption{8.8~$M_{\odot}$
progenitor mass. Radial evolution of the ratio
$R$ between the matter $\lambda$ and neutrino  $\mu$ potentials at different
post-bounce times.} \label{fig1}
\end{figure}
%%%%%%%%%%%%%%%%%%%%%%%%%%%%%%%%%%%%%%%%%%%%%%%%%%%%%%%%%%%%%%%%%%%%%%%%%%%%%%%%%%%%%%%%%%%%%%%%%%%%%%%%%%%%%%%%%%  

In this section we analyze the flavor conversions for the model of $8.8$~$M_{\odot}$ O-Ne-Mg progenitor.
In Fig.~4 we plot the ratio $R=\lambda/\mu$ for the same time snapshots of Fig.~9 of~\cite{Chakraborty:2011gd},
i.e. in the range   [0.08,0.25]~s. 
We realize that in this case there is no abrupt discontinuity associated with the shock front. 
Indeed, for this low-mass progenitor 
there is no extended accretion phase, since  the explosion succeeds
very shortly after the core-bounce. Therefore, the shock-front is already beyond the radial range interesting for
the flavor conversions.
Since in this case
the matter density of the envelope is  low compared
to the iron-core progenitors, the electron density profile
above the core is very steep. Therefore, 
at $r \gtrsim$ few hundred km, 
the  ratio  $R \lesssim 3$ [for $t_{\rm pb} \in [0.08,0.16]~s$ it monotonically decreases
becoming also smaller than 1] suggests that flavor conversions could arise there.

In Fig.~5 we show the radial evolution of  the eigenvalue $\kappa$ for the time snapshots shown in Fig.~4,  determined 
from the solution of Eq.~(\ref{consit}). We use the same format of Fig.~2. 
In Fig.~6 we show the corresponding survival probability $P_{ee}$ for electron antineutrinos ${\bar\nu}_e$.
Starting with the case without matter term, i.e. $\lambda=0$ (dashed curves), 
we see that the $\kappa$ function rapidly becomes larger than 1, and the peak corresponds to the 
onset of the flavor conversions in Fig.~6.  
We pass now considering the case with $\lambda$ and half-isotropic neutrino angular 
distributions (continuous curves). 
We realize that, for $t_{\rm pb}= 0.08, 0.25$~s in the region where $\kappa$ grows, 
$R\simeq 1-2$. Therefore,  the matter suppression of the instability is never  complete. Moreover, the rise of $\kappa$ is rapid and the position of the peak corresponds
to the onset of the flavor conversions seen in Fig.~6.
Conversely, for the other time snapshots ($t_{\rm pb}= 0.14, 0.16$~s) where $R \gtrsim 3$, the suppression of 
the instability is stronger. Moreover, the $\kappa$ curves are broadened and there is not a clear peak. 
Therefore, one cannot easily link a non-zero $\kappa$ with the numerical onset of the flavor conversions. 
In the case of  $\lambda$ and flavor dependent forward-peaked neutrino angular
distributions (dotted curves), 
as expected we find a  stronger suppression in the flavor instability. 
However, as shown in Fig.~1 of ~\cite{Saviano:2012yh}, 
the angular spectra of different flavors
for the
 8.8 M$_{\odot}$ SN are significantly
less forward-peaked than in the case of the  10.8 M$_{\odot}$ SN.
Therefore, their effect is less pronounced in   this case. In particular, for
$t_{\rm pb}= 0.8, 0.25$~s, $\kappa$ is  large enough  to trigger flavor conversions.
Conversely, these  are strongly inhibited at $t_{\rm pb}= 0.14$~s, and completely suppressed
for $t_{\rm pb}= 0.16$~s. 

Finally, we checked also in this case that including a possible halo effect does not change the results of the stability analysis.
 In conclusion, for our  SN model with 8.8 M$_{\odot}$ progenitor mass the matter suppression of 
 the MAA instability is not complete at early times. Therefore, in principle one would expect interesting time-dependent 
 features in the observable neutrino spectra.

%%%%%%%%%%%%%%%%%%%%%%%%%%%%%%%%%%%%%%%%%%%%%%%%%%%%%%%%%%%%%%%%%%%%%%%%%%%%%%%%%%%%%%%%%%%%%%%%%%%%%%%%%%%%%
\begin{figure}[!t]
\begin{center}
 \includegraphics[angle=0,width=0.5\textwidth]{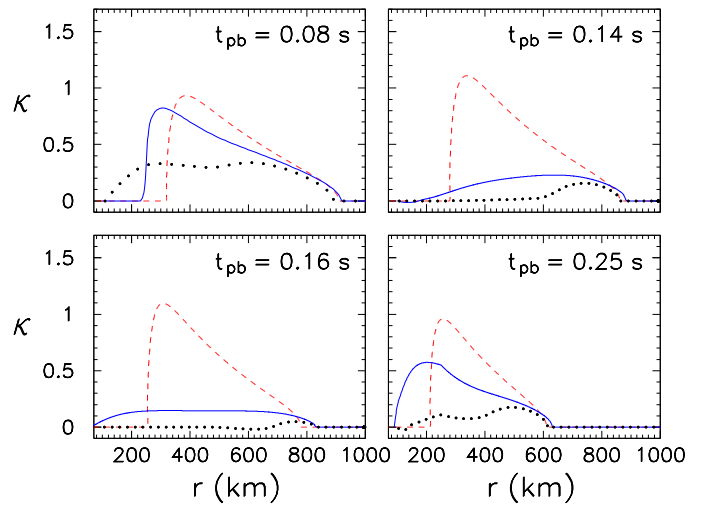} 
    \end{center}
\caption{8.8 M$_{\odot}$ progenitor mass. Radial evolution of the 
$\kappa$ function (in units of km$^{-1}$) at different post-bounce times
 with $\lambda=0$ for a half-isotropic neutrino emission
(dashed curves)  and in presence of matter effects, with 
a half-isotropic neutrino emission (continuous curves) and 
with flavor-dependent angular distributions (dotted curves).} \label{fig5}
\end{figure}
%%%%%%%%%%%%%%%%%%%%%%%%%%%%%%%%%%%%%%%%%%%%%%%%%%%%%%%%%%%%%%%%%%%%%%%%%%%%%%%%%%%%%%%%%%%%%%%%%%%%%%%%%%%%%%%%%%  

%%%%%%%%%%%%%%%%%%%%%%%%%%%%%%%%%%%%%%%%%%%%%%%%%%%%%%%%%%%%%%%%%%%%%%%%%%%%%%%%%%%%%%%%%%%%%%%%%%%%%%%%%%%%%
\begin{figure}[!t]
\begin{center}
 \includegraphics[angle=0,width=0.5\textwidth]{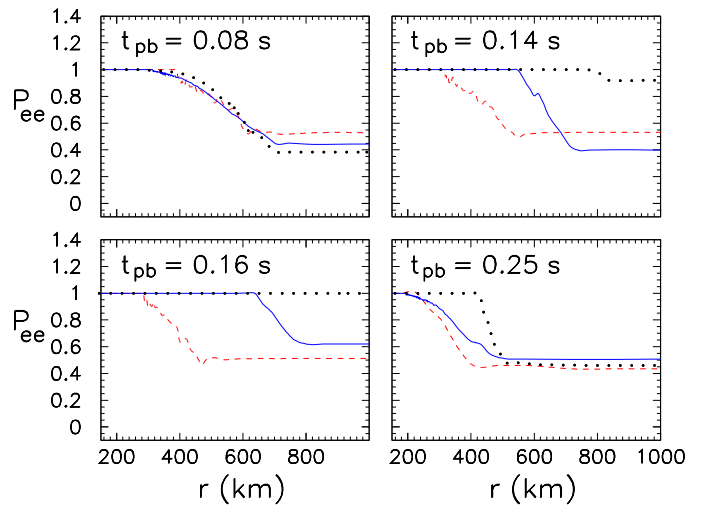} 
    \end{center}
\caption{8.8 M$_{\odot}$ progenitor mass. Radial evolution of the
survival probability $P_{ee}$
for electron antineutrinos at different
post-bounce times for the MAA evolution with $\lambda=0$ for a half-isotropic neutrino emission
(dashed curves)  and in presence of matter effects, with 
a half-isotropic neutrino emission (continuous curves) and 
with flavor-dependent angular distributions (dotted curves).} \label{fig6}
\end{figure}
%%%%%%%%%%%%%%%%%%%%%%%%%%%%%%%%%%%%%%%%%%%%%%%%%%%%%%%%%%%%%%%%%%%%%%%%%%%%%%%%%%%%%%%%%%%%%%%%%%%%%%%%%%%%%%%%%% 

 %%%%%%%%%%%%%%%%%%%%%%%%%%%%%%%%%%%%%%%%%%%%%%%%%%%%%%%%%%%%%%%%%%%%%%%%%%%%%55
 \section{Conclusions}
 %%%%%%%%%%%%%%%%%%%%%%%%%%%%%%%%%%%%%%%%%%%%%%%%%%%%%%%%%%%%%%%%%%%%%%%%%%%%%%%55
 
 We have performed  a dedicated study  of the matter suppression  of the
MAA instability, connected with the axial symmetry breaking in the self-induced oscillations,  during the accretion phase   for two SN models with different progenitor masses. We characterize  the SN densities and
 the neutrino angular spectra with results from recent SN hydrodynamical simulations.
We compared the linear stability analysis  with the numerical results of the flavor evolution  in which we 
have looked  at a local solution of the equations of motions along a specific line of sight.
For the case of an iron-core 10.8 M$_{\odot}$  we found that during the accretion phase the dominant matter
density strongly suppresses the MAA instability. In particular, including
realistic forward-peaked $\nu$ angular distributions significantly
reduces the strength of the $\nu$-$\nu$ interaction term. As a result, in this case the matter suppression of the 
self-induced flavor conversions would be complete. 
In the case of  a low-mass O-Ne-Mg SN with 8.8 M$_{\odot}$ progenitor, where  the
 accretion phase is extremely short, 
the matter density profile is lower  and the $\nu$ angular distributions less forward-peaked 
than in iron-core models. As a consequence, 
we found that also with realistic angular distributions flavor conversions   would be    possible at early times,
producing in principle interesting time-dependent modulations. 

Our analysis is complementary to  previous studies~\cite{Chakraborty:2011nf,Chakraborty:2011gd,Saviano:2012yh}, where some of us explored the matter suppression of self-induced flavor
conversions in inverted neutrino mass hierarchy, in axial-symmetric models
 (see also~\cite{Sarikas:2011am}). 
The complete suppression of the self-induced effects in both the mass  hierarchies for iron-core SNe, 
 implies that the neutrino signal during the accretion phase will be processed only by the ordinary 
Mikheyev-Smirnov-Wolfenstein   effect in the outer stellar layers.
This effect would allow in principle  to distinguish the neutrino mass hierarchy through the  
  study of the  rise time of the SN 
neutrino signal~\cite{Serpico:2011ir}.
The phenomenological importance of our findings motivates further studies with other SN models
to confirm the generality of our results. In particular, an accurate characterization of the neutrino angular
distributions seems necessary in order to get accurate predictions on the matter suppression. 
At this regard,  in~\cite{Sawyer:2008zs} it has been shown
that if  perfect cylindrical symmetry in the initial neutrino distributions
were given up, then  super-fast flavor turnovers could occur. 
These effects cannot be tested within our spherically symmetric SN model. 
However, recently three-dimensional SN simulations have been carried on, 
characterizing the neutrino signal during the accretion phase. Surprisingly,
a lepton-emission asymmetry among different flavor has been found~\cite{Tamborra:2014aua}. 
In particular, the  electron (anti)neutrino fluxes show a dipole structure, while
the $\nu_x$ are almost spherically symmetric. We plan to investigate in a future
work the role of the matter suppression in this flavor configuration, including also
the not axisymmetric neutrino and matter angular distributions. 

Self-induced flavor conversions associated with the MAA instability would still be possible 
for O-Ne-Mg SNe during the accretion phase, and possibly for iron core SNe  during the cooling phase, when the matter term becomes sub-dominant with respect to the neutrino-neutrino interaction
term. 
In these situations, a self-consistent treatment of the neutrino equations of motion considering 
also the flavor evolution along the  transverse
direction is still lacking. This would imply passing from an ordinary to a partial differential equation problem, adding a big
layer of complication in the solution of the equations of motion. 
This effort is well motivated by the perspective of getting an accurate  characterization of the SN neutrino
spectral features that would be observable in the planned large underground neutrino detectors~\cite{Choubey:2010up}.

%%%%%%%%%%%%%%%%%%%%%%%%%%%%%%%%%%%%%%%%%%%%%%%%%%%%%%%%%%%%%%%%%%%%%%
\section*{Acknowledgements} %%%%%%%%%%%%%%%%%%%%%%%%%%%%%%%%%%%%%%%%%%%%%%%%
%%%%%%%%%%%%%%%%%%%%%%%%%%%%%%%%%%%%%%%%%%%%%%%%%%%%%%%%%%%%%%%%%%%%%%

We thank T.~Fischer, G.~Raffelt, S.~Sarikas and M.~Wu for useful discussions.
S.C.\ acknowledges support from the European Union through a Marie Curie Fellowship, Grant No.\ PIIF-GA-2011-299861, and through the ITN ``Invisibles'', Grant No.\ PITN-GA-2011-289442.
The work of A.M.  was supported by the German Science Foundation (DFG)
within the Collaborative Research Center 676 ``Particles, Strings and the
Early Universe.'' 
N.S. acknowledges support from
the European Union FP7 ITN INVISIBLES (Marie Curie
Actions, PITN- GA-2011- 289442).
D.S. acknowledges support by the Funda\c{c}\~{a}o
para a Ci\^{e}ncia e Tecnologia (Portugal) under grant
SFRH/BD/66264/2009.

%%%%%%%%%%%%%%%%%%%%%%%%%%%%%%%%%%%%%%%%%%%%%%%%%%%%%%%%%%%%%%%%%%%%%%
%% References %%%%%%%%%%%%%%%%%%%%%%%%%%%%%%%%%%%%%%%%%%%%%%%%%%%%%%%%
%%%%%%%%%%%%%%%%%%%%%%%%%%%%%%%%%%%%%%%%%%%%%%%%%%%%%%%%%%%%%%%%%%%%%%

\end{document}